\documentclass[aps,prc,twocolumn,showpacs,floatfix,nofootinbib,preprintnumbers,superscriptaddress,amsmath,amssymb,longbibliography,amsproc]{revtex4-1}
\usepackage{epsfig,dsfont,amssymb,amsmath,amsthm,amsfonts,amsbsy,mathrsfs}
\usepackage{graphicx}
\usepackage{amsmath}
\usepackage{amssymb}%
\usepackage{dcolumn}
\usepackage{hyperref}

\usepackage[disable]{todonotes}

\newcommand{\be}{\begin{equation}}
\newcommand{\ee}{\end{equation}}
\newcommand{\ba}{\begin{eqnarray}}
\newcommand{\ea}{\end{eqnarray}}

\begin{document}

\title{Charge radii of exotic neon and magnesium isotopes}

\author{S.~J.~Novario}
\affiliation{Department of Physics and Astronomy, University of Tennessee, Knoxville, TN 37996, USA}
\affiliation{Physics Division, Oak Ridge National Laboratory, Oak Ridge, TN 37831, USA}

\author{G.~Hagen}
\affiliation{Physics Division, Oak Ridge National Laboratory, Oak Ridge, TN 37831, USA}
\affiliation{Department of Physics and Astronomy, University of Tennessee, Knoxville, TN 37996, USA}
\affiliation{TRIUMF, 4004 Wesbrook Mall, Vancouver BC, V6T 2A3, Canada}

\author{G.~R.~Jansen}
\affiliation{National Center for Computational Sciences, Oak Ridge National Laboratory, Oak Ridge, Tennessee 37831, USA}
\affiliation{Physics Division, Oak Ridge National Laboratory, Oak Ridge, TN 37831, USA}

\author{T.~Papenbrock}
\affiliation{Department of Physics and Astronomy, University of Tennessee, Knoxville, TN 37996, USA}
\affiliation{Physics Division, Oak Ridge National Laboratory, Oak Ridge, TN 37831, USA}

\begin{abstract}
We compute the charge radii of even-mass neon and magnesium isotopes
from neutron number $N=8$ to the dripline. Our calculations are based
on nucleon-nucleon and three-nucleon potentials from chiral effective
field theory that include delta isobars. These potentials yield an
accurate saturation point and symmetry energy of nuclear matter. We
use the coupled-cluster method and start from an axially symmetric
reference state. Binding energies and two-neutron separation energies
largely agree with data and the dripline in neon is accurate. The
computed charge radii have an estimated uncertainty of about 2-3\% and
are accurate for many isotopes where data exist. Finer details such as
isotope shifts, however, are not accurately reproduced. Chiral
potentials correctly yield the subshell closure at $N=14$ and also a
decrease in charge radii at $N=8$ (observed in neon and predicted for
magnesium). They yield a continued increase of charge radii as
neutrons are added beyond $N=14$ yet underestimate the large increase
at $N=20$ in magnesium.
\end{abstract}

%\pacs{21.30.-x, 21.30.Fe, 21.10.Dr, 21.60.-n}

\maketitle

{\it Introduction. ---} The radii of atomic nuclei carry information
about their structure as isotopic trends reflect changes in nuclear
deformation, shell structure, superconductivity (pairing), and weak
binding.  The difference between the radii of the neutron and proton
distributions of an atomic nucleus also impact the structure of
neutron stars. Matter radii are usually extracted from reactions with
strongly interacting probes, which requires a model-dependent
analysis~\cite{tanihata1985,tanihata2013}.  In contrast, electric
charge radii (and more recently also weak charge radii) can be
determined using the precisely known electroweak
interaction~\cite{abrahamyan2012,hagen2015}. Precision measurements of
nuclear charge radii have contributed much to our understanding of
stable nuclei and rare isotopes, and they continue to challenge
nuclear structure
theory~\cite{bissell2016,garciaruiz2016,miller2019,groote2020}.

In the past two decades we have seen a lot of progress in ab initio
computations of nuclei, i.e. calculations that employ only controlled
approximations and are based on Hamiltonians that link the nuclear
many-body problem to the nucleon-nucleon and few-nucleon
systems. Virtually exact
methods~\cite{pieper2001,navratil2009,barrett2013,carlson2015} scale
exponentially with increasing mass number and depend on the
exponential increase of available computational cycles for progress.
A game changer has been combining ideas and soft interactions from
effective field theory
(EFT)~\cite{vankolck1994,entem2003,epelbaum2009,machleidt2011,hebeler2011,ekstrom2015}
and the renormalization group~\cite{bogner2003,bogner2007,bogner2010},
with approximate (but systematically improvable) approaches that scale
polynomially with mass number. Examples of such methods are
coupled-cluster theory~\cite{mihaila2000b,dean2004,hagen2014},
in-medium similarity renormalization
group~\cite{tsukiyama2011,hergert2016}, nuclear lattice
EFT~\cite{lee2009,lahde2014}, and self-consistent Green's function
approaches~\cite{dickhoff2004,soma2013}.

Nuclei as heavy as $^{100}$Sn have now been computed within this
framework~\cite{morris2018}, and the first survey of nuclei up to mass
50 or so has appeared~\cite{holt2019}. Computing nuclei is much
more costly than using, e.g., nuclear density-functional theory
(DFT)~\cite{bender2003,niksic2011,erler2012}. However, the
ever-increasing availability of computational cycles makes these
computations both feasible and increasingly affordable.  The approach
based on Hamiltonians offers the possibility to compute excited
states, to perform symmetry projections, and to treat currents and
Hamiltonians consistently. It migh also be possible to link such
Hamiltonians back to quantum
chromodynamics~\cite{barnea2015,contessi2017,mcilroy2018,bansal2018}.

\begin{figure}
  \includegraphics[width=0.53\textwidth]{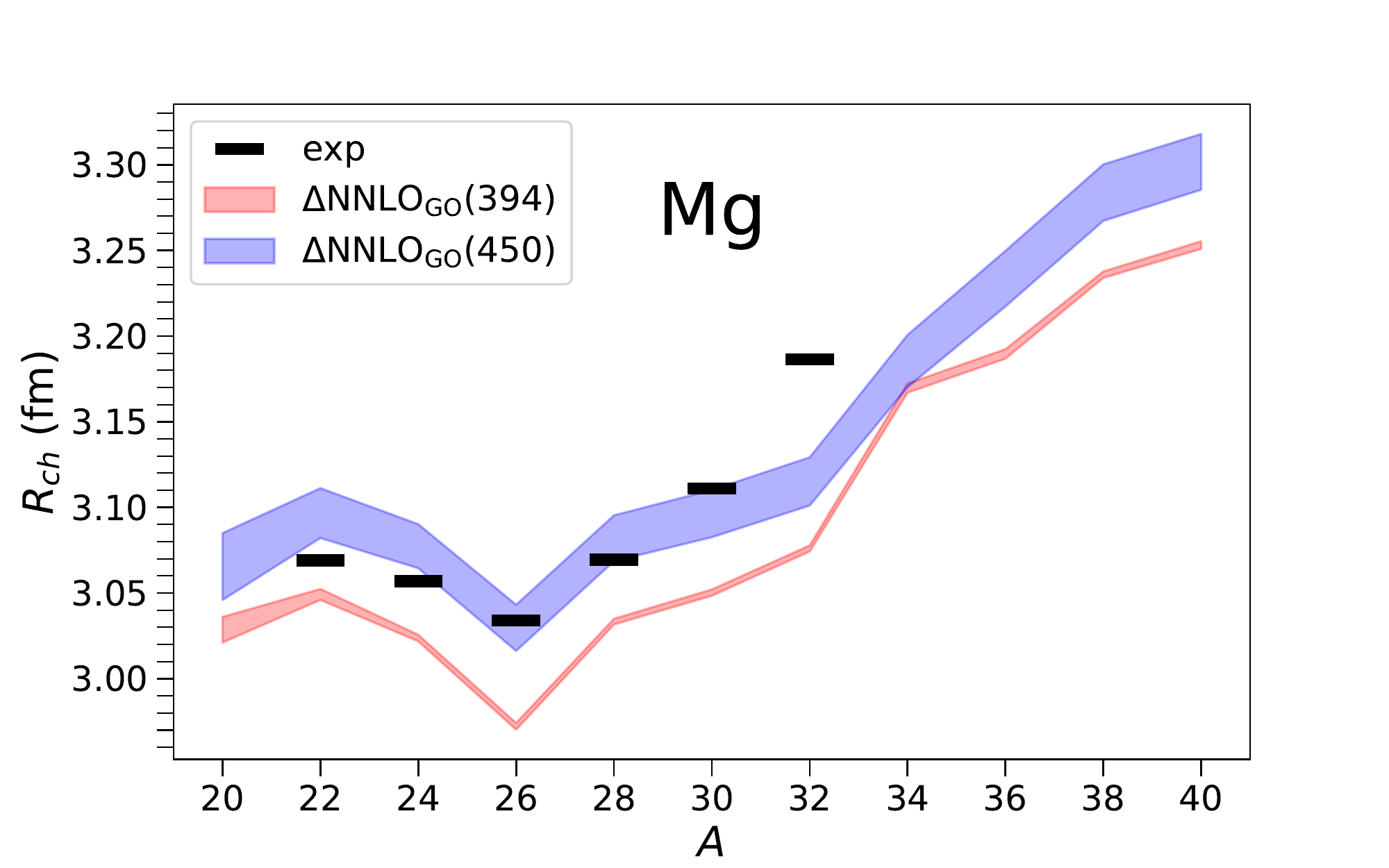}
  \caption{(Color online) Charge radii for magnesium isotopes with
    even mass numbers computed with the potentials $\Delta$NNLO$_{\rm
      GO}(394)$ (red) and $\Delta$NNLO$_{\rm GO}(450)$ (blue) compared
    to data (solid bars)~\cite{yordanov2012}. The model spaces consist
    of 13 oscillator shells with oscillator frequencies
    $\hbar\omega=12$ and 16~MeV, as indicated by the bands.
      \label{fig:Mg-radii}}
\end{figure}

In this paper, we compute the charge radii of neutron-rich isotopes of
neon and magnesium. These nuclei are at the center of the island of
inversion~\cite{poves1987,warburton1990} and at the focus of current
experimental interests. Charge radii are known for
$^{17-28}$Ne~\cite{marinova2011,angeli2013} and
$^{21-32}$Mg~\cite{yordanov2012} (see Fig.~\ref{fig:Mg-radii}),
leaving much to explore. Of particular interest is the impact (or lack
thereof) of the ``magic'' neutron numbers $N=8,14$, and 20 on charge
radii, the onset of deformation past $N=20$, and the rotational
structure of neutron-rich isotopes as the dripline is
approached~\cite{crawford2019,vilen2020}.

Among the many available interactions from chiral
EFT~\cite{epelbaum2009,machleidt2011,hebeler2011,ekstrom2013,entem2015,epelbaum2015,huether2019,soma2020}, NNLO$_{\rm
  sat}$~\cite{ekstrom2015} and $\Delta$NNLO$_{\rm
  GO}$~\cite{payne2019,bagschi2020,jiang2020} stand out through their
quality in describing nuclear radii. These interactions contain pion
physics, three-nucleon forces, and -- in the case of
$\Delta$NNLO$_{\rm GO}$ -- effects of the $\Delta$ isobars. Both
interactions have been constrained by data on the nucleon-nucleon
interaction, and nuclei with mass numbers $A=3,4$. While NNLO$_{\rm
  sat}$ also was constrained by binding energies and radii of nuclei
as heavy as oxygen, $\Delta$NNLO$_{\rm GO}$ was constrained by the
binding energy, density and symmetry energy of nuclear matter at its
saturation point. These potentials use the leading-order three-body
forces from chiral EFT~\cite{hebeler2015b}. In this study we will
employ two $\Delta$NNLO$_{\rm GO}$ interactions which differ by their
respective momentum cutoffs of 394 and 450~MeV$c^{-1}$.

\begin{figure}[htb]
  \includegraphics[width=0.48\textwidth]{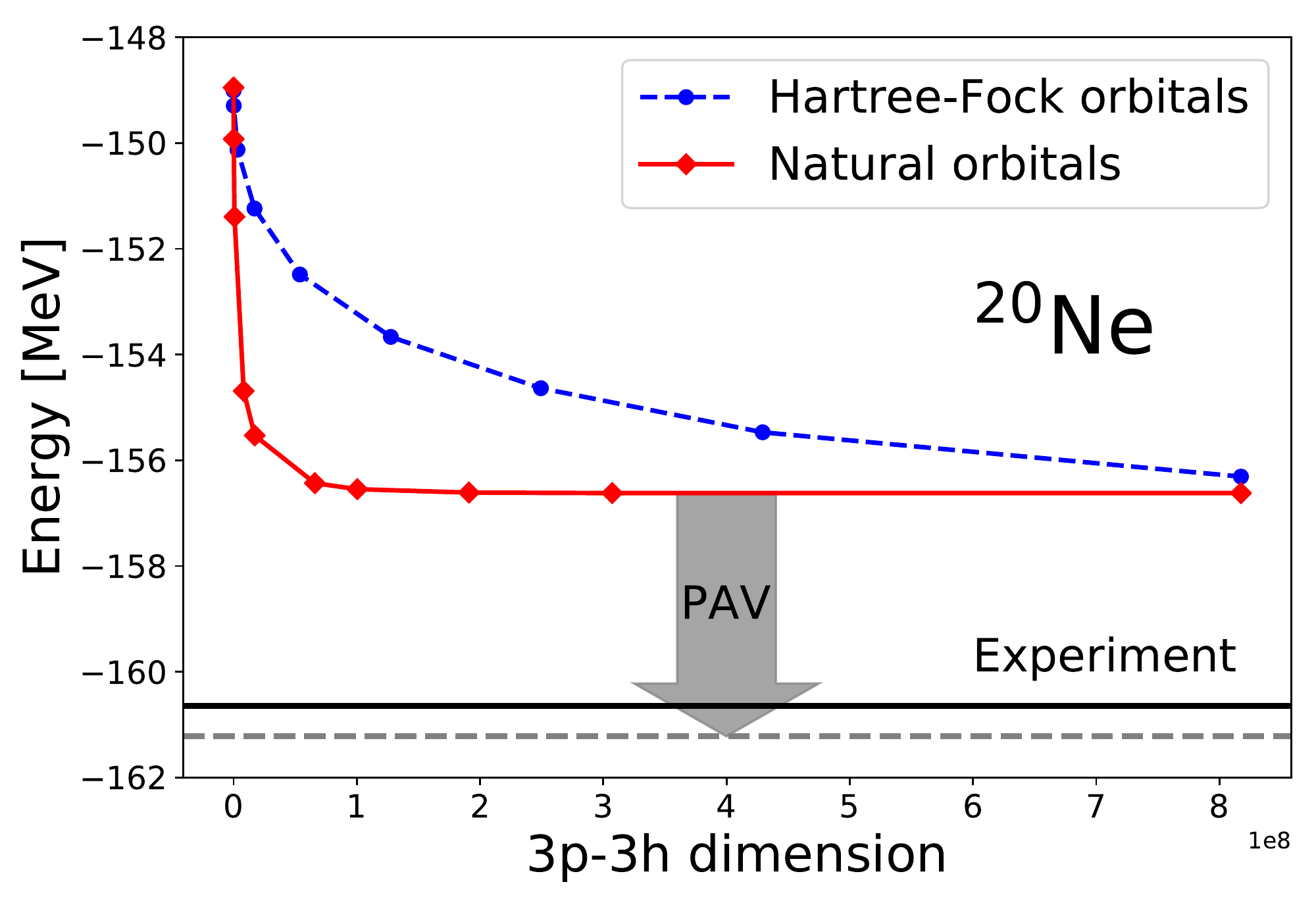}
  \caption{(Color online) Ground state energy of $^{20}$Ne with
    respect to the number of included $3p$--$3h$ amplitudes, computed from
    the Hartree-Fock basis (blue circles connected by dashed line),
    and natural orbitals (red diamonds connected by full line).  For
    the Hartree-Fock basis we limited the number of $3p$--$3h$ excitations
    by the energy cut
    $\tilde{E}_{pqr}=\tilde{e}_p+\tilde{e}_q+\tilde{e}_r<\tilde{E}_{\rm
      3max}$, where $ \tilde{e}_p=|N_p-N_{F}|$ is the energy
    difference between the single-particle energies and the Fermi
    surface $N_F$. The cut in the natural orbital basis is described
    in the main text.  We used the $\Delta$NNLO$_{\rm GO}(394)$
    potential and a model-space of 11 major spherical oscillator
    shells with the frequency $\hbar\omega = 16$~MeV.  The black solid
    line is the experimental value, while the gray dashed line
    includes the energy gain from projection after variation of the
    Hartree-Fock result.
    \label{fig:Ne-nat}}
\end{figure}

{\it Theoretical approach.--- } Our coupled-cluster calculations
start from an axially deformed product state built from natural
orbitals. To construct the natural orbitals we perform a Hartree-Fock
calculation that keeps axial symmetry, parity, and time-reversal
symmetry, but is allowed to break rotational invariance.  Thus, the
$J_z$ component of angular momentum is conserved, and single-particle
orbitals come in Kramer-degenerate pairs with $\pm j_z$. For
open-shell nuclei, we fill the partially occupied neutron and proton
shells at the Fermi surface from low to high values of $|j_z|$; this
creates a prolate Hartree-Fock reference. Following
Ref.~\cite{tichai2019} we use this state to compute its density matrix
in second-order perturbation theory and diagonalize it to obtain the
natural orbitals. As shown in Fig.~\ref{fig:Ne-nat} and discussed
below, natural orbitals improve the convergence of the ground-state
energies with respect to the number of three-particle--three-hole
($3p$--$3h$ amplitudes in the coupled-cluster wave-function.

The natural orbital basis is spanned by up to 13~spherical harmonic
oscillator shells. We present results for two different oscillator
frequencies ($\hbar\omega=12$ and 16~MeV) to gauge the model-space
dependence. The three-nucleon interaction had the additional energy
cut of $E_\mathrm{3max}=16 \hbar\omega$, which is sufficient to
converge the energies and radii reported in this work.

The breaking of rotational symmetry by the reference state is
consistent with the emergent symmetry breaking and captures the
correct structure of the nontrivial
vacuum~\cite{yannouleas1999}. However, our approach lacks possible
tri-axial deformation and symmetry restoration, for which several
proposals
exist~\cite{duguet2015,qiu2017,qiu2018,tsuchimochi2018}. Overcoming
these limitations is thus possible but comes at a significant increase
in computational cost: The loss of symmetries (either by permitting
tri-axiality or by rotating the Hamiltonian during projection)
significantly increases the number of non-zero Hamiltonian matrix
elements and coupled-cluster amplitudes. To estimate the impact of
symmetry restoration we performed projection after variation of the
deformed Hartree-Fock states for all nuclei considered in this work,
and found an energy gain from 3 to 6~MeV. This provides us with an
upper limit on the energy that can be gained through symmetry
restoration, as we would expect that correlations beyond the
mean-field partially restore broken symmetries. We note that tri-axial
deformations in the ground-state are not expected to be significant
for the nuclei we study in this paper~\cite{moller2006}.  We finally
note that the axially-symmetric coupled-cluster computations are an
order of magnitude more expensive than those that keep rotational
invariance. Fortunately, the availability of leadership-class
computing facilities and the use of graphics processor units (GPUs)
now make such computations possible.

Our calculations start from the ``bare'' Hamiltonian
\be
H = T_{\rm kin} -T_{\rm CoM} +V_{NN} +V_{NNN}
\ee
based on the $\Delta$NNLO$_{\rm GO}$ nucleon-nucleon and three-nucleon
potentials $V_{NN}$ and $V_{NNN}$, respectively. Here, $T_{\rm kin}$
denotes the kinetic energy, and we subtract the kinetic energy of the
center of mass $T_{\rm CoM}$ to remove the center-of-mass from the
Hamiltonian. We express this Hamiltonian in terms of operators
$\hat{a}^\dagger_p$ and $\hat{a}_q$ that create and annihilate a
nucleon with quantum numbers $q$ and $p$, respectively, in the natural
orbital basis. The Hamiltonian $H_N$ is normal-ordered with respect to
the reference state, and we only retain up to normal-ordered two-body
forces; we have $H_N=F_N + V_N$, where the Fock term $F_N$ denotes the
normal-ordered one-body part and $V_N$ the normal-ordered two-body
terms~\cite{shavittbartlett2009}.

The coupled-cluster
method~\cite{coester1958,kuemmel1978,bishop1991,mihaila2000b,bartlett2007,dean2004,binder2013,hagen2014}
generates a similarity-transformed Hamiltonian
\ba
\label{Hbar}
\overline{H}_N\equiv e^{-\hat{T}}H_N e^{\hat{T}} \ ,
\ea
using the cluster-excitation operator
\ba
\label{cc}
\hat{T} &=& \hat{T}_1 +\hat{T}_2 +\hat{T}_3 \cdots \nonumber\\
&=&\sum_{ia}t_i^a \hat{a}_a^\dagger \hat{a}_i + {1\over 4}\sum_{ijab}t_{ij}^{ab}\hat{a}_a^\dagger \hat{a}_b^\dagger \hat{a}_j\hat{a}_i \nonumber\\
&&+{1\over 36}\sum_{ijkabc}t_{ijk}^{abc}\hat{a}_a^\dagger \hat{a}_b^\dagger \hat{a}_c^\dagger \hat{a}_k\hat{a}_j\hat{a}_i + \cdots .
\ea
The operator $\hat{T}_n$ creates $n$-particle--$n$-hole excitations of
the reference state $|\psi\rangle \equiv \prod_{i=1}^A
\hat{a}_i^\dagger|0\rangle$. Here and in what follows, labels $i,j,k$ refer to
single-particle states occupied in the reference state, while $a,b,c$
are for unoccupied states.

We truncate the expansion~(\ref{cc}) at the $3p$--$3h$ level and include
leading-order triples using the CCSDT-1
approximation~\cite{lee1984,watts1995}. In this approximation $e^T
\approx e^{T_1 +T_2} +T_3$, and the amplitudes $t_i^a$, $t_{ij}^{ab}$,
and $t_{ijk}^{abc}$ fulfill
\ba
\label{ccsdt1}
\langle \psi_i^a|\overline{H}_N + H_N\hat{T}_3 |\psi\rangle &=&0 \ , \nonumber\\
\langle \psi_{ij}^{ab}|\overline{H}_N + H_N\hat{T}_3 |\psi\rangle &=&0 \ , \nonumber\\
\langle \psi_{ijk}^{abc}|\left(F_N\hat{T}_3 +V_N\hat{T}_2\right)_{\rm con}|\psi\rangle &=&0 \ .
\ea
In the first two lines $\hat{T}=\hat{T}_1+\hat{T}_2$ enters the
similarity transformation, which gives the commonly used
coupled-cluster singles-and-doubles (CCSD) approximation when $T_3 =
0$. In the last line only the connected terms enter. The correlation
energy is then $E_0=\langle\psi|\overline{H}_N|\psi\rangle$.

The CCSD approximation costs $o^2u^4$ compute cycles for each
iteration, with $o=A$ ($u$) being the number of (un)occupied states
with respect to the natural-orbital reference. The cost
of CCSDT-1 is $o^3u^4$ and thus an order of magnitude more
expensive.

Both CCSD and CCSDT-1 are too expensive without further
optimizations. To overcome this challenge we first take advantage of
the block-diagonal structure of the Hamiltonian imposed by axial
symmetry, isospin, and parity and only store and process
matrix-elements that obey these symmetries. Second, we impose a
truncation on the allowed number of $3p$--$3h$ amplitudes by a cut on the
product occupation probabilities $n_a$ for three particles above the
Fermi surface and for three holes below the Fermi surface, i.e.  we
require $ n_a n_b n_c \leq \varepsilon$ and
$(1-n_i )(1-n_j) (1-n_k) \leq \varepsilon$.  This cut favors
configurations with large occupation probabilities near the Fermi
surface and -- as shown in Fig.~\ref{fig:Ne-nat} -- requires only a
manageable number of $3p$--$3h$ amplitudes to be included.
Third, we exploit the internal structure of the three-body symmetry
blocks, which can be expressed as the tensor product of two- and
one-body symmetry blocks, to formulate the equations as a series of
matrix multiplications. This allows us to efficiently utilize the
supercomputer Summit at the Oak Ridge Leadership Computing Facility,
whose computational power mainly comes from GPUs.

For the computation of observables other than the energy (the radius
in our case), we also need to solve the left eigenvalue problem as the
similarity transformed Hamiltonian is non-Hermitian. This is done
using the equation-of-motion coupled-cluster method (EOM-CCM), see
Refs.~\cite{stanton1993, watts1995, bartlett2007,hagen2014} for
details.  In this work we limit the computations of radii to the
EOM-CCSD approximation level. For $^{32}$Mg the inclusion of
(computationally expensive) triples via the EOM-CCSDT-1
approximation~\cite{watts1995} increases the radius by less than 1\%,
consistent with the findings of
Refs.~\cite{miorelli2018,kaufmann2020}.

In EOM-CCSDT-1 the left ground-state eigenvalue problem is
\begin{equation}
\label{left}
\langle \psi|  (1+\hat{\Lambda}) \overline{H}_N  =E_0 \langle\psi| (1+\hat{\Lambda})\ .
\end{equation}
Here $\Lambda $ is a de-excitation operator with amplitudes $\Lambda_a^i$,
$\Lambda_{ab}^{ij}$, and $\Lambda_{abc}^{ijk}$.
We need to solve for
\ba
\hat{\Lambda} &=&\hat{\Lambda}_1 + \hat{\Lambda}_2 +\hat{\Lambda}_3 \nonumber\\
&=&\sum_{ia}\Lambda^i_a \hat{a}_i^\dagger \hat{a}_a + {1\over 4}\sum_{ijab}\Lambda^{ij}_{ab}
\hat{a}_i^\dagger \hat{a}_j^\dagger \hat{a}_b\hat{a}_a \nonumber\\
&&+{1\over 36}\sum_{ijkabc}\Lambda^{ijk}_{abc}\hat{a}_i^\dagger \hat{a}_j^\dagger \hat{a}_k^\dagger \hat{a}_c\hat{a}_b\hat{a}_a
\ea
Given $\overline{H}_N$, Eq.~(\ref{left}) is an eigenvalue problem, and
we are only interested in its ground-state solution $E=E_0$. In the
EOM-CCSDT-1 approximation, the triples de-excitation part $\Lambda_3$
only contributes to the doubles de-excitation part of the
matrix-vector product via $\langle \psi \vert \hat{\Lambda}_3 V_N
\vert \psi^{ij}_{ab}\rangle$, while the triples de-excitation part of
the matrix-vector product is $\langle \psi \vert (\hat{\Lambda}_1 +
\hat{\Lambda}_2 )V_N + (\hat{\Lambda}_2 + \hat{\Lambda}_3 )F_N \vert
\psi^{ijk}_{abc}\rangle$.  To compute the left ground-state we can
either solve a large-scale linear problem (because we know the
ground-state energy $E_0$), or we use an iterative Arnoldi algorithm
for general non-symmetric eigenvalue problems to compute the ground
state of $\overline{H}_N$.  In our experience the latter approach is
more stable and requires fewer iterations. The ground-state
expectation value of an operator $\hat{O}$ is
\be
\langle \hat{O}\rangle \equiv \langle\psi|(1+\hat{\Lambda})\overline{O}|\psi\rangle \ .
\ee
Here the similarity-transformation $\overline{O}\equiv
e^{-\hat{T}}\hat{O}e^{\hat{T}}$ of $\hat{O}$ enters.

The charge radius squared is
\be
R_{\rm ch}^2 = R_p^2 +\langle r_p^2\rangle +{N\over Z}\langle r_n^2\rangle +\langle r_{\rm DF}^2\rangle +\langle r_{\rm SO}^2\rangle \ .
\ee
Here, $R_p^2$ is the radius squared of the intrinsic point-proton
distribution and $\langle r_{\rm SO}^2\rangle$ is the spin-orbit
corrections. These two quantities are actually computed with the
coupled-cluster method~\cite{hagen2015}. The corrections
$\langle r_p^2\rangle= 0.709$~fm$^2$,
$\langle r_n^2\rangle=-0.106$~fm$^2$, and
$\langle r_{\rm DF}^2\rangle=3/(4m^2)=0.033$~fm$^2$ (with $m$ denoting
the nucleon mass) are the charge radius squared of the proton (updated
according to Refs.~\cite{pohl2010,xiong2019}), the neutron (updated
value from Ref.~\cite{filin2020}), and the Darwin-Foldy term,
respectively.

{\it Results.---} Our results for the charge radii of magnesium
isotopes are shown in Fig.~\ref{fig:Mg-radii}. Here, each band
reflects model-space uncertainties from varying the oscillator
frequency from 12 to 16~MeV. The results for the softer interaction
with a cutoff of 394~MeV$c^{-1}$ are shown in red and exhibit less
model-space dependence than those for the harder interaction with
450~MeV$c^{-1}$ shown in blue.  The overall uncertainty estimate on
the radii, both from model-space uncertainties and systematic
uncertainties of the interactions is then about 2-3\%, i.e. the full
area covered by (and between) both bands.

Overall, the $\Delta$NNLO$_{\rm GO}$ potentials reproduce the
prominent pattern of a minimum radius at the sub-shell closure $N=14$,
and they agree with data within uncertainties for mass numbers $22\le
A\le 30$. The computed radii continue to increase beyond $N=14$, and
they reflect the absence of the $N=20$ shell closure in
magnesium. This is, of course, the beginning of the island of
inversion. However, the theory results do not reproduce the very steep
increase from $A=30$ to 32. Thus, they seem to reflect remnants of a
shell closure at $N=20$ that are not seen in the data.  Theory
predicts increasing charge radii as the dripline is approached.  This
is consistent with an increase in nuclear deformation as neutrons are
added~\cite{crawford2019}. We also note that theory predicts a marked
shell closure at $N=8$ for neutron-deficient magnesium. This is in
contrast to the trend projected in Ref.~\cite{yordanov2012}. The
excited $2^+$ state in $^{18}$Mg at 1.6~MeV is somewhat higher that
the 1.2~MeV observed in $^{20}$Mg, and the question regarding a
sub-shell closure at $N=8$ is thus undecided. It will be interesting
to compare the theoretical results with upcoming laser spectroscopy
experiments that are at the proposal stage~\cite{vilen2020}.

\begin{figure}[htb]
  \includegraphics[width=0.53\textwidth]{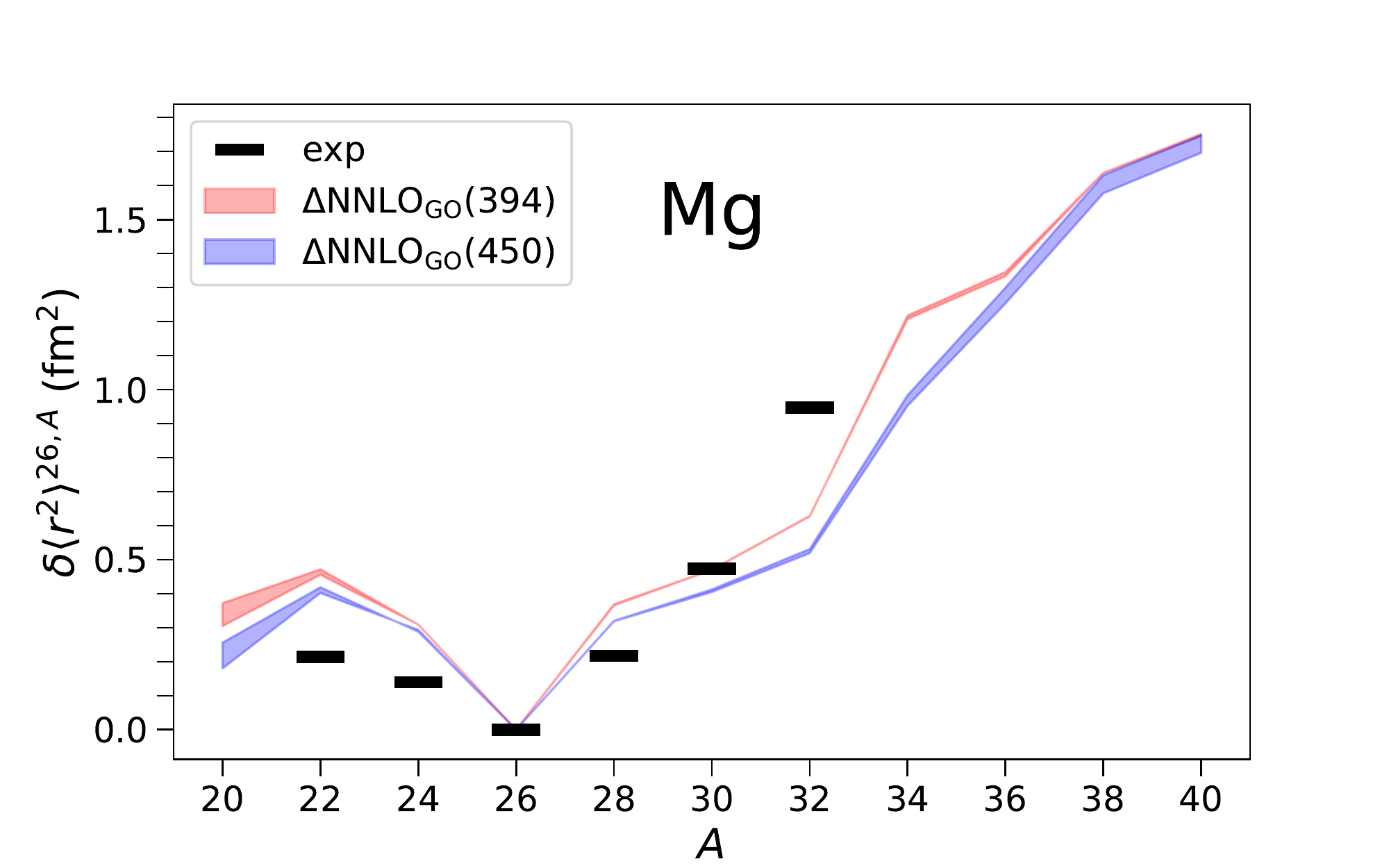}
  \caption{(Color online) As in Fig.~\ref{fig:Mg-radii} but for the
    isotope shift, i.e. the charge radii squared relative to $^{26}$Mg.
  \label{fig:Mg-delR}}
\end{figure}

The plot of isotopic variations in the charge radii, shown in
Fig.~\ref{fig:Mg-delR}, is interesting. Theory is not accurate
regarding most isotopes shifts and over-emphasizes shell closures at
$N=14$ and $N=20$ that are not in the data.  This is perhaps a most
important result of this study: While state-of-the-art potentials can
now describe charge radii within 2-3\% of relative uncertainties,
finer details such as isotope shifts still escape the computations.

\begin{figure}[htb]
  \includegraphics[width=0.53\textwidth]{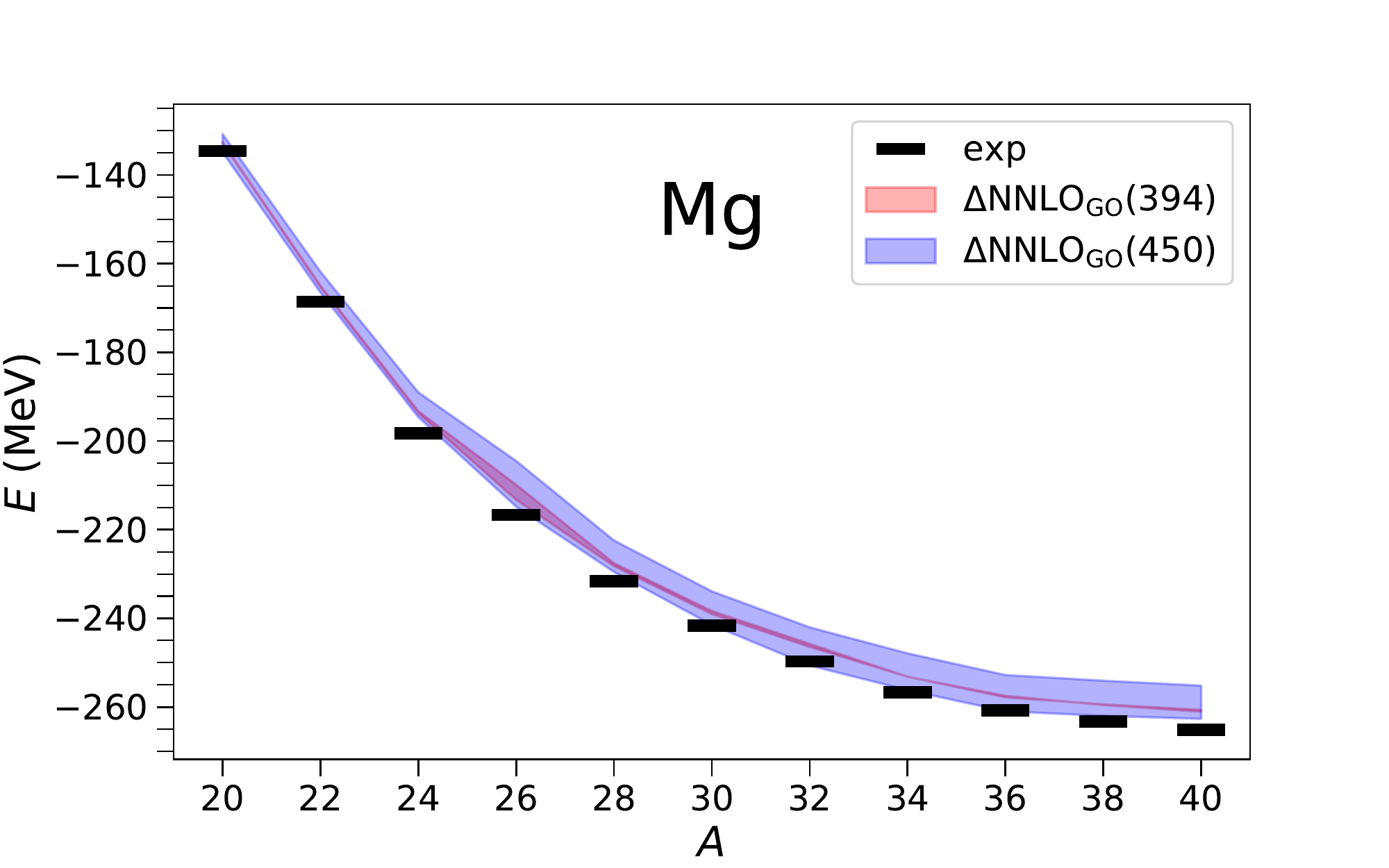}
  \caption{(Color online) As in Fig.~\ref{fig:Mg-radii} but for the
    ground-state energies.
  \label{fig:Mg-Egs}}
\end{figure}

We show the results for binding energies in Fig.~\ref{fig:Mg-Egs}. Our
calculations yield the dripline at $^{40}$Mg, with $^{42}$Mg being
about 1.8~MeV less bound for the $\Delta$NNLO$_{\rm GO}(394)$
potential. However, computational limitations prevented us from
including continuum effects, which can easily yield an additional
binding energy of the order of 1~MeV~\cite{hagen2016}. This prevents
us from predicting the unknown dripline in magnesium more
precisely~\cite{baumann2007}.

Another uncertainty stems from the lack of angular-momentum
projection.  To estimate the corresponding energy correction, we
performed a projection after variation within the Hartree-Fock
computations. These projections lower the Hartree-Fock energy by about
3 to 5~MeV, see Fig.~\ref{fig:Ne-nat} for an example. We expect that a
projection of the coupled-cluster results would yield slightly less
energy gains (because these calculations already include some of the
correlations that are associated with a projection). Overall,
Fig.~\ref{fig:Mg-Egs} shows that the $\Delta$NNLO$_{\rm GO}$
potentials accurately describe nuclear binding energies also for
open-shell nuclei.

\begin{figure}[htb]
  \includegraphics[width=0.53\textwidth]{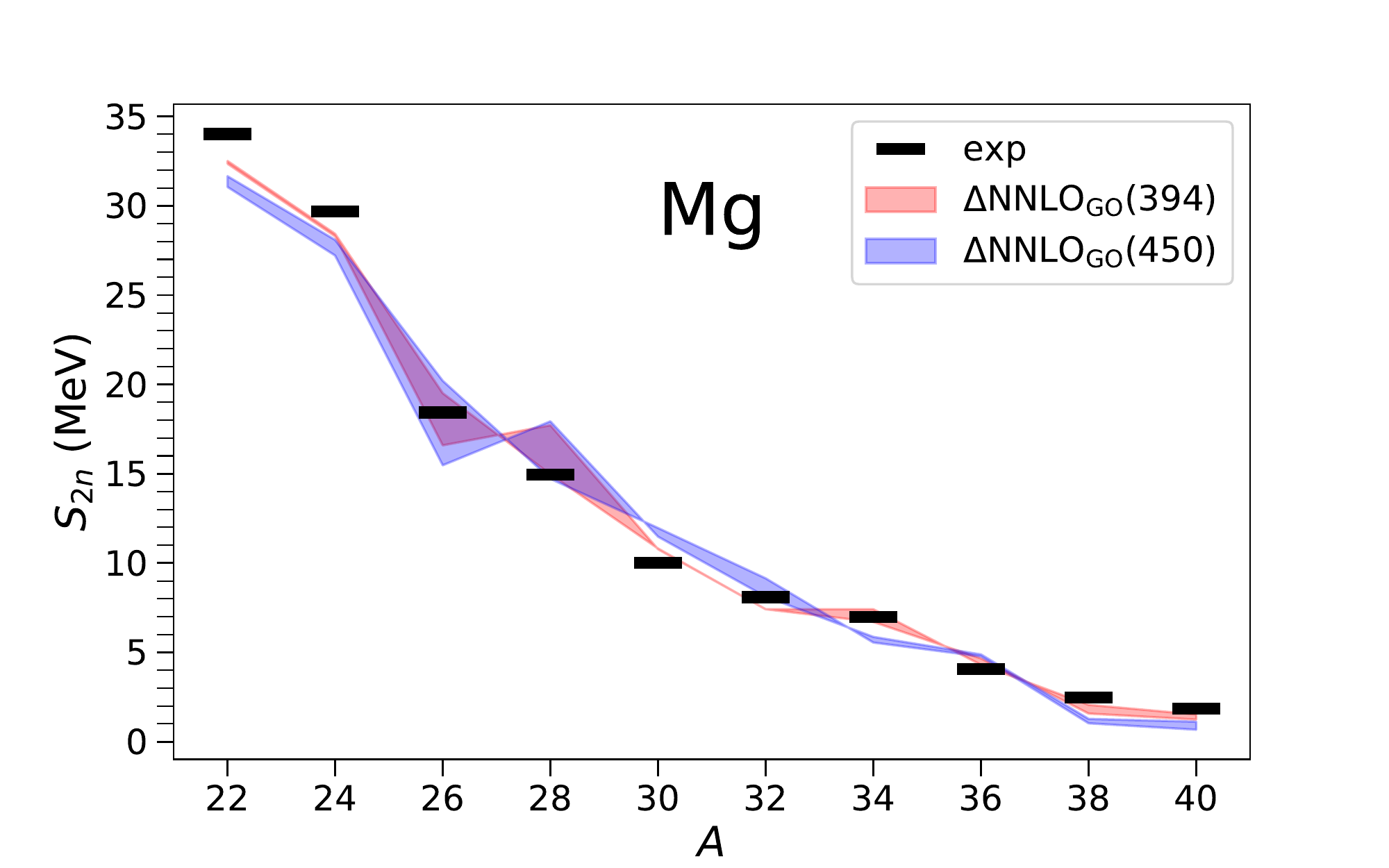}
  \caption{(Color online) As in Fig.~\ref{fig:Mg-radii} but for the
    two-neutron separation energies
  \label{fig:Mg-S2n}}
\end{figure}

Binding-energy differences, such as the two-neutron separation energy,
is another observable sensitive to shell structure and dripline
physics.  Figure~\ref{fig:Mg-S2n} shows that the overall pattern in
the data is accurately reproduced within the uncertainties from the
employed interactions and model spaces. However, the details of the
sub-shell closure at $N=14$ escape the theoretical description,
i.e. theory predicts a slightly stronger sub-shell than observed
experimentally.

\begin{figure}[htb]
  \includegraphics[width=0.53\textwidth]{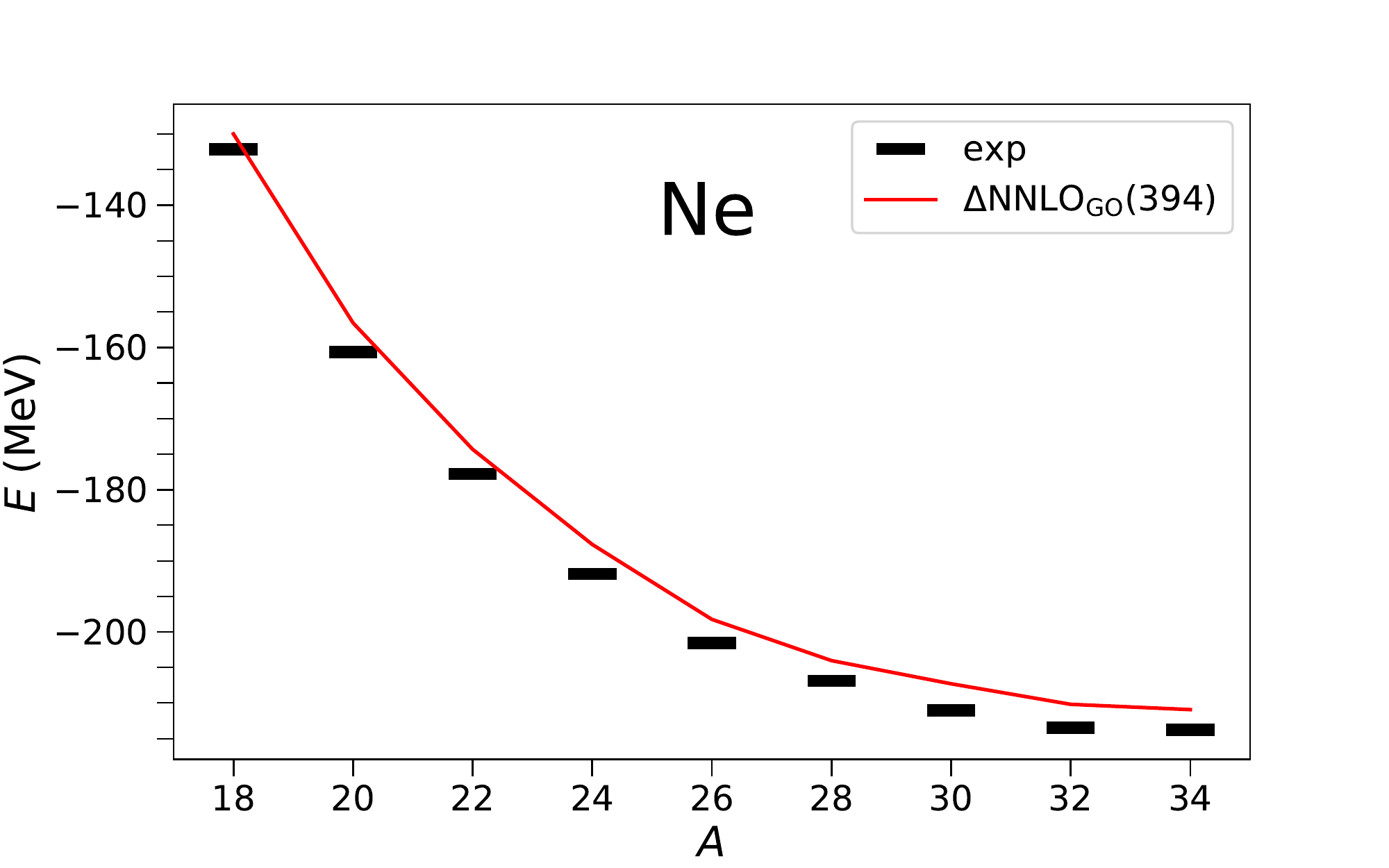}
  \caption{(Color online) Ground-state energies for neon isotopes with
    even mass numbers computed with the potentials $\Delta$NNLO$_{\rm
      GO}(394)$ shown as a red line. The model spaces consist of 13
    oscillator shells. Data is shown as black bars.
  \label{fig:Ne-Egs}}
\end{figure}

We finally turn to neon isotopes. Here, our computations have been
less extensive to manage the available computational cycles.  We
limited the computations of energies to the $\Delta$NNLO$_{\rm
  GO}(394)$ potentials in a model space of 13 harmonic oscillator
shells at $\hbar\omega=16$~MeV.  For the charge radii we also employed
the $\Delta$NNLO$_{\rm GO}(450)$ potential at $\hbar\omega=12$~MeV.
Figure~\ref{fig:Ne-Egs} shows that the ground-state energies are close
to the data. We estimate theoretical uncertainties to be a bit smaller
than for the magnesium isotopes. We also note that about 3-5~MeV of
energy gain is expected from a projection of angular momentum (see
again Fig.~\ref{fig:Ne-nat}). We find the dripline at $^{34}$Ne, in
agreement with data~\cite{ahn2019}.

\begin{figure}[htb]
  \includegraphics[width=0.53\textwidth]{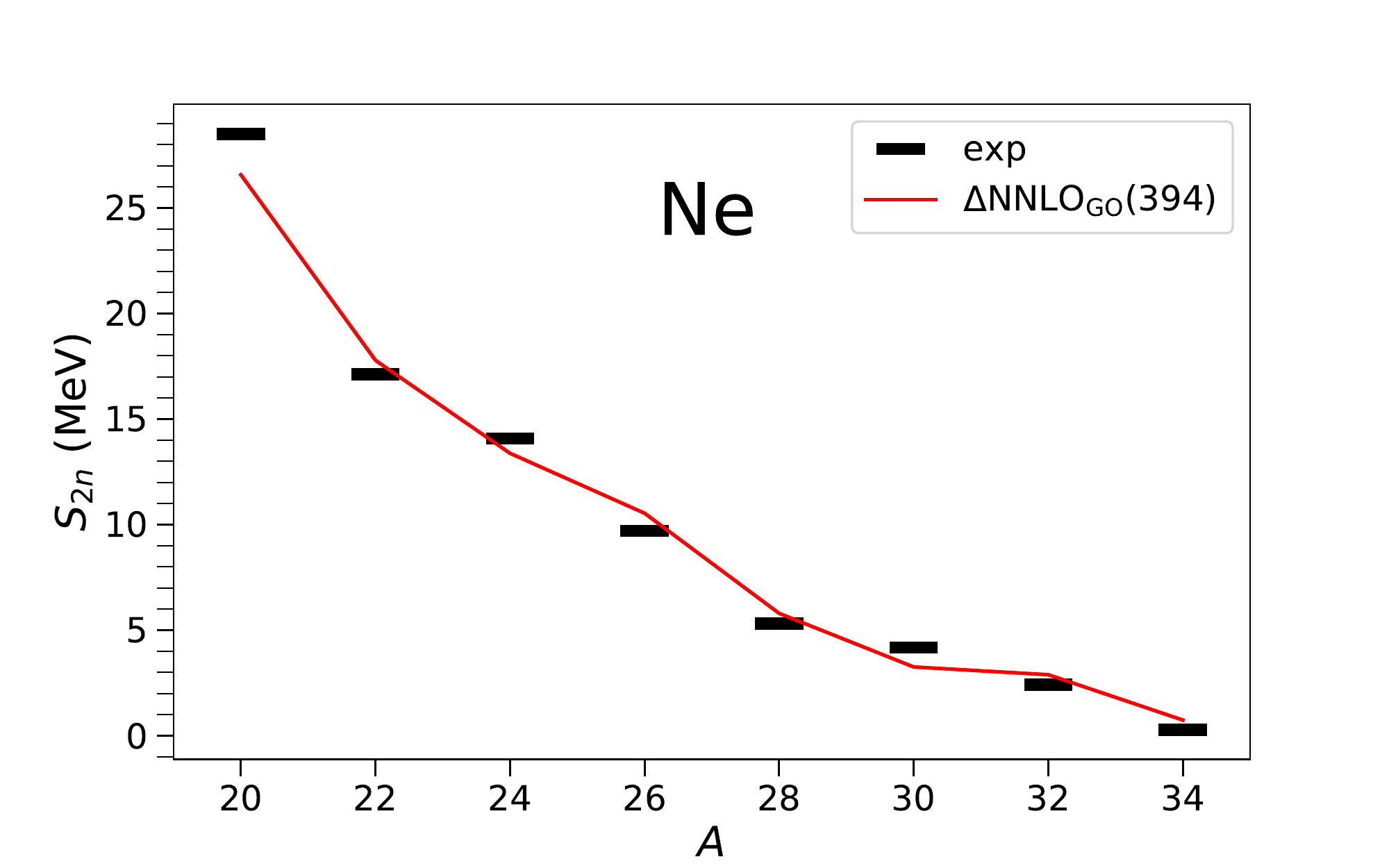}
  \caption{(Color online) As in Fig.~\ref{fig:Ne-Egs} but for the
    two-neutron separation energies.
  \label{fig:Ne-S2n}}
\end{figure}

The computed two-neutron separation energies, shown in
Fig.~\ref{fig:Ne-S2n}, confirm this picture. Compared to magnesium, it
is interesting that the addition of two protons shifts the drip line
by about six neutrons. Again we estimate that theoretical
uncertainties are a bit smaller than for the magnesium isotopes.

\begin{figure}[htb]
  \includegraphics[width=0.53\textwidth]{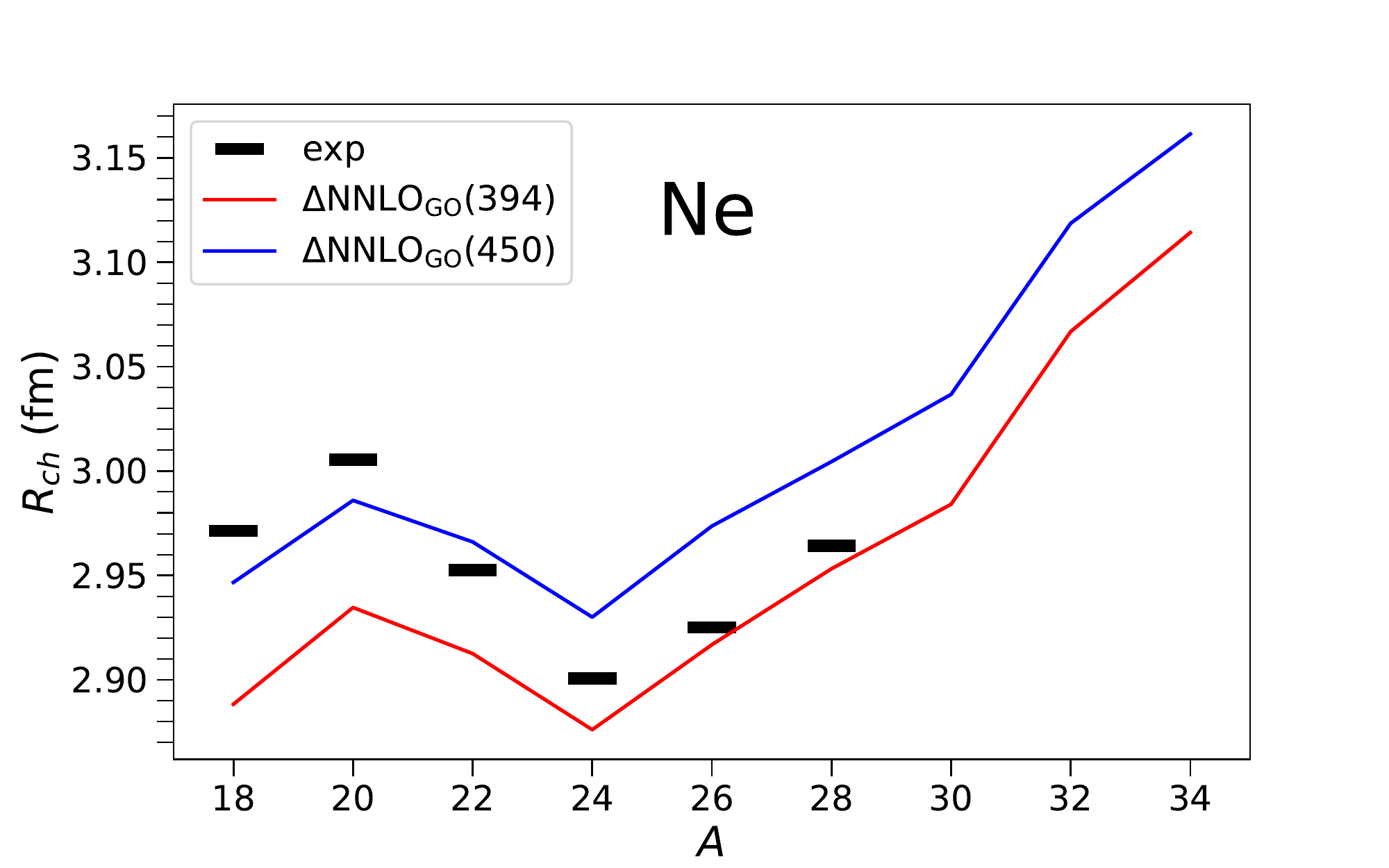}
  \caption{(Color online) Charge radii for neon isotopes with even
    mass numbers computed with the potentials $\Delta$NNLO$_{\rm
      GO}(394)$ and $\Delta$NNLO$_{\rm GO}(450)$ shown in red and
    blue, respectively. The model spaces consist of 13 oscillator
    shells. Data is shown as black
    bars~\cite{marinova2011,angeli2013}.
  \label{fig:Ne-Rch}}
\end{figure}

Finally, we show results for charge radii in Fig.~\ref{fig:Ne-Rch},
using the $\Delta$NNLO$_{\rm GO}(394)$ and $\Delta$NNLO$_{\rm
  GO}(450)$ potentials. We only employed one oscillator frequency for
each interaction. Thus, the theoretical uncertainties are estimated to
be somewhat larger than the area between the two lines (compare with
Fig.~\ref{fig:Mg-radii} of the magnesium isotopes). Based on these
estimates, theoretical results are not quite accurate below $^{22}$Ne,
though they qualitatively reproduce the overall trend. The results
accurately reflect the known sub-shell closures at $N=14$ and
$N=8$. We see no closure at $N=20$ and it will be interesting to
confront this prediction with data.

{\it Conclusion.---} We computed ground-state energies, two-neutron
separation energies, and charge radii for neon and magnesium
isotopes. Our computations were based on nucleon-nucleon and
three-nucleon potentials from chiral EFT, and we employed
coupled-cluster methods that started from an axially symmetric
reference state. The computed energies and radii are accurate when
taking expected corrections from angular momentum projection into
account. Trends in charge radii, and the minimum and neutron number
$N=14$ are qualitatively reproduced. Within our estimated
uncertainties of about 2-3\%, however, quantitative accuracy is not
achieved for all isotopes, and isotope shifts still challenge
theory. Nevertheless, we predict a continuous increase as the neutron
dripline is approached, and this is consistent with a considerable
nuclear deformation. Proposed experiments will soon confront these
predictions.

\begin{acknowledgments}
We thank T. Duguet, Z. H. Sun, and A. Tichai for useful discussions.
This work was supported by the U.S. Department of Energy, Office of
Science, Office of Nuclear Physics, under Award Nos.~DE-FG02-96ER40963
and DE-SC0018223.  Computer time was provided by the Innovative and
Novel Computational Impact on Theory and Experiment (INCITE)
programme. This research used resources of the Oak Ridge Leadership
Computing Facility located at Oak Ridge National Laboratory, which is
supported by the Office of Science of the Department of Energy under
contract No. DE-AC05-00OR22725.

\end{acknowledgments}

%\bibliography{master,refs}
%\clearpage % flush figures before references

\end{document}